\documentclass[%
 reprint,
 amsmath,amssymb,
 aps,
prd,
]{revtex4-2}

\usepackage{graphicx}
\usepackage{dcolumn}
\usepackage{bm}
\usepackage{hyperref}
\usepackage[mathlines]{lineno}
\usepackage{cleveref}
\usepackage{booktabs}
\usepackage{mathtools}

\begin{document}

\title{Jump-Diffusion Stochastic Quantization for Euclidean Lattice Field Theories}

\author{Alexander Rothkopf}
\affiliation{Department of Physics, Korea University, Seoul 02841, Republic of Korea}
\email{akrothkopf@korea.ac.kr}

\date{\today}

\begin{abstract}
We construct the natural generalization of stochastic quantization (in the Markovian sense) by considering jump-diffusion processes. This class of stochastic processes exhibits non-continuous paths, so-called L\'evy flights. In the presence of jumps, action landscapes with barriers can be efficiently explored, improving and even restoring ergodicity where traditional diffusion approaches become inefficient. We explore different strategies for constructing efficient jump updates, which we deploy to address the benchmark problem of topological freezing in 2d U(1) gauge theory.
\end{abstract}


\maketitle


\section{Motivation}

Almost any question we ask of a Euclidean quantum field theory can be cast in the form of computing averages
\begin{align}
    \langle O \rangle = \frac{1}{Z}\int {\cal D}\phi\, O(\phi) \,e^{-S[\phi]}, \quad Z=\int {\cal D}\phi\, e^{-S[\phi]}.
\end{align}
Here $S[\phi]$ denotes the action $S=\int d^{d+1}x {\cal L}_E[\phi]$ formulated in Euclidean time $x^0=\tau$.
Discretized on a $(d+1)$ dimensional space-time lattice these averages represent highly dimensional integrals. The most common approach for their evaluation is importance sampling, where one constructs a stochastic process that visits field configurations $\phi$ with frequency proportional to ${\rm exp[-S[\phi]]}$ and averages realizations of observables $O$ over those configurations. 

In conventional stochastic quantization \cite{Parisi:1980ys,Damgaard:1987rr} this stochastic process is a diffusion process. It is described by a stochastic partial differential equation formulated in an artificial Langevin time $t$, distinct from the Euclidean time $\tau$ of the action
\begin{align}
    \partial_{t} \phi(t,x) = -\frac{\delta S[\phi]}{\delta \phi(t,x)} + \sqrt{2} \eta(t,x).
\end{align}
Here $\eta$ represents independent Gaussian white noise sources. As a diffusion process governed by a drift and noise term, the field moves in infinitesimal continuous steps.

Diffusion processes as samplers have two known failure modes \cite{Sokal:1996cargese}: \textbf{\textit{ (i) Barriers and first order transitions}}: If ${\rm exp[-S[\phi]]}$ exhibits several well separated peaks \cite{Palmer:1982be} coresponding to distinct minima of the action. These arise from e.g. phase coexistence or the presence of distinct topological sectors. in that case the continuous path must climb the barrier, which happens with a probability ${\rm exp[-\Delta S_{\rm barrier}]}$ (for a detailed analysis see e.g. \cite{Kramers:1940,Hanggi:1990rmp}). I.e. the stochastic process gets trapped and can look perfectly converged, while sampling only a single sector. One way to address this problem is to flatten the barrier due to a modification of the system and then to reweight to the original ensemble, as explored in tempering \cite{Marinari:1992qd,Hukushima:1996,Earl:2005pt} and multicanonical approaches \cite{Berg:1991multicanonical,Janke:1998multicanonical,WangLandau:2001}. \textbf{\textit{ (ii) Critical slowing down and frustrated systems}}: Near a continuous phase transition, correlated regions grow in size and it takes increasing time $\tau\sim\xi^z$, associated with the correlation length $\xi$ of the system \cite{Hohenberg:1977ym}, to update them by local moves \cite{Swendsen:1987ce}. Similarly, in frustrated systems \cite{Houdayer:2001}, local updates are unable to find acceptable configurations. Only by constructing collective updates \cite{Fortuin:1971dw,Wolff:1988uh,Brower:1989mt,WangLandau:2001} over extended regimes of the simulation domain can efficient sampling be recovered.

In this study we propose to cure the barrier failure mode by extending stochastic quantization from diffusion to jump-diffusion (for textbooks see e.g. \cite{del2017stochastic,Gardiner:2004,Applebaum:2009} and for a classic application in finance \cite{merton1976option}). I.e. we will allow the fields to make occasional finite moves, on top of their diffusive behavior. These jumps are constructed in order to sample the target distribution ${\rm exp[-S[\phi]]}$.

One area of research where the loss of ergodicity due to barriers plays an important role is the numerical simulation of nuclear matter, known as lattice QCD. The gluon gauge fields carry a topological charge and changes between charge sectors incur a finite action cost. Sampling close to the continuum limit, one finds that samplers which appear highly efficient for local observables are unable to cross the barriers between topological sectors. For a characterization of the challenge see e.g. Ref.~\cite{DelDebbio:2002xa} and Ref.~\cite{Schaefer:2010hu}.

A wealth of methods has been proposed over the past two decades to improve the sampling of topological sectors. We may accept the fact that freezing happens and attempt to correct for it a posteriori as in Refs.~\cite{Brower:2003yx,Aoki:2007ka}. Alternatively one may consider geometric strategies, such as sampling on very large grids \cite{Luscher:2017cjh} or introducing open boundary conditions \cite{Luscher:2011kk}. One may also consider introducing a bias in the system \cite{Laio:2015era,Eichhorn:2023uge} or deploy a multicanonical approach \cite{Jahn:2018dke} to encourage transitions between adjacent sectors or combinations thereof \cite{Bonanno:2020hht}.

Recent interest has focussed on the concept of trivializing maps \cite{Luscher:2009eq} and their use in generative machine learning models, so-called normalizing flows (see e.g. Refs.~\cite{Albergo:2019eim,Kanwar:2020xzo,Bonanno:2025flow} ), as well as diffusion models \cite{Wang:2023exq,Fukushima:2024oij,Aarts:2026zzr} to achieve sampling with reduced autocorrelations.

Considering jumps to restore ergodicity is not a new idea per se. It has been explored in the guise of modified Metropolis-Hastings (MH) and Hybrid Monte Carlo (HMC) updates in various studies in the past. A recent study in Ref.~\cite{Albandea:2021lvq} e.g. amends the continuous trajectory updates of the HMC with non-continuous changes to the field configuration, called windings of definite topological charge. When accepted these induce changes between topological sectors. Earlier work in this direction, known as instanton hits is discussed in Refs.~\cite{Fucito:1983qg,Dilger:1994ma}. 

In this work we construct the framework of jump-diffusion stochastic quantization, in which jumps are the natural generalization of continuous path diffusion updates (which include deterministic evolution). Update strategies, such as windings and instanton hits, naturally find their place inside this approach. Importantly, the insight we gain from the general treatment of the jump structure will allow us to design update strategies that improve on the efficiency of known jump extensions.

With the presence of jumps, stochastic quantization can also be applied to systems with discrete degrees of freedom, to which the concept of diffusion does not apply. It is from the research community on such discrete systems that we take inspiration to construct efficient jump updates in the following sections.

The structure of the paper is as follows. In \cref{sec:jdsq} we develop the theoretical framework of jump-diffusion stochastic quantization, focussing on the simplest class of jumps, so-called Poisson jump of constant rate in \cref{sec:pjsq}. After establishing the detailed balance condition needed for sampling of the correct target distribution, we study three different strategies to design jumps in \cref{sec:jumpd}. In \cref{sec:analyzej}, following similar analyses in the HMC literature, we derive an estimate for acceptance revealing which quantity must be optimized so that our jump updates are efficient. \Cref{sec:application} applies the jump-diffusion framework to sampling to simple model systems. In \cref{sec:tdw} ergodicity is restored in the tilted double well before we consider two-dimensional U(1) gauge theory in \cref{sec:2dU1}. Here we first explore an optimal jump strategy based on extended configurations \cref{sec:globu}, before working with local updates \cref{sec:locu}, which are needed for transferring the jump strategy to more realistic systems in the future. \Cref{sec:conclo} summarizes the results of the paper and provides an outlook on future research direction with jump-diffusion stochastic quantization.

\section{Jump-diffusion stochastic quantization}
\label{sec:jdsq}
\subsection{Markov processes}
\label{sec:conclo}
According to classic theorems of probability theory, Courr\`ege's theorem for the generator of a stochastic process \cite{Courrege:1965} and the L\'evy-It\^o decomposition for the stochastic paths \cite{Applebaum:2009,del2017stochastic}, the most general
continuous-time Markov process consists of exactly three ingredients. These are deterministic drift, Gaussian diffusion, and jumps. Starting from the Kramers-Moyal expansion, the Taylor expansion for the master equation describing the evolution of a probability distribution $P(\phi,\tau_L)$ under a general Markov process, one finds that the time evolution can be cast in the form of the differential Chapman-Kolmogorov (CK) equation
\begin{widetext}
\begin{align}
\partial_t P(\phi,t)=
-\partial_\phi\big[b(\phi)P\big] + \partial_\phi\partial_\phi\big[aP\big]
+\int d \phi'\Big[W(\phi|\phi')P(\phi')-W(\phi'|\phi)P(\phi)\Big],
\label{eq:generalfp}
\end{align}
\end{widetext}
The first two terms are the well known contributions to the continuous diffusive dynamics described by the ordinary Fokker-Planck equation. The third term encodes the discontinuous jump contributions with an intuitive gain-loss structure for jumps towards and away from the current configuration. The fact that the last term is not local is a consequence of Pawula's theorem \cite{Pawula:1967} that requires that if the Kramers-Moyal expansion goes beyond the second order term, an infinite number of higher order terms are contributing. 

For diffusion stochastic quantization without kernel the drift term $b=-\delta S/\delta \phi$ and noise term $a=1$. A kernel $K$ modifies $b\to Kb$ and $a\to\sqrt{K}a$. This work constructs appropriate transition rate densities $W(x|x')$ for the jumps. Within the context of time continuous Markov processes adding jumps completes the framework and no further ingredients remain.

\subsection{Poisson jumps}
\label{sec:pjsq}

As the simplest jump-diffusion process let us consider jumps occurring with constant rate $\lambda_0$ on top of a diffusion process
\begin{align}
    d\phi(t) = b(\phi)dt +\sqrt{2}dW_t,
\end{align}
where $dW_t$ is the increment of a Wiener process with $\mathbb{E}[dW]=0$ and $\mathbb{E}[dW^2]=dt$.  

Following the conventional strategy laid out in textbooks \cite{del2017stochastic,Applebaum:2009,Gardiner:2004} let us construct the Chapman-Kolmogorov equation for Poisson jump-diffusion in detail, which offers a concrete realization of the theorems mentioned above. In a single infinitesimal time step the process at hand with probability $1-\lambda_0 dt$ simply diffuses or with $\lambda_0 dt$ executes a jump. In case that a jump occurs the field is displaced by a finite value $\eta$ drawn from the jump distribution $q(\eta|\phi)$. Since jumping and not jumping are mutually exclusive we have
\begin{widetext}
\begin{align}
\mathbb E\big[f(\phi_{t+dt})\big]
&=(1-\lambda_0 dt)\;\mathbb E\big[f(\phi+b dt+\sqrt2\,dW)\big]
+\lambda_0 dt\;\mathbb E\!\left[\int d\eta\,q(\eta|\phi)\,f(\phi+\eta)
\right]+\mathcal O(dt^2)
\nonumber\\
&=\mathbb E\big[f(\phi)\big]
+dt\,\mathbb E\Big[\underbracket{b f'(\phi)+f''(\phi)}_{\text{It\^o
expansion}}
+\lambda_0\!\int d\eta\,q(\eta|\phi)\big[f(\phi+\eta)-f(\phi)\big]\Big]
+\mathcal O(dt^2),
\end{align}
\end{widetext}
The second line involves the It\^o-Taylor expansion of $f$ to second order. In addition we have absorbed the term $-dt \lambda_0 \mathbb{E}[f(\phi)]$ into the jump integral exploiting the normalization $\int d\eta q(\eta|\phi)=1$. We conclude that the evolution of the expectation value can be summarized by a generator $L$ 
\begin{align}
&\frac{d}{dt}\,\mathbb E[f]=\mathbb E\big[(Lf)(\phi)\big]\\
\nonumber&(Lf)(\phi)=\\
\nonumber &b(\phi)f'(\phi)+f''(\phi)
+\lambda_0\!\int d\eta\;q(\eta|\phi)\big[f(\phi+\eta)-f(\phi)\big],
\label{eq:genexplicit}
\end{align}
which exhibits the Courr\`ege form \cite{Courrege:1965}. Now to obtain the CK equation we must relate the expectation value to the underlying probability distribution via $\mathbb E[f]=\int d\phi\,P(\phi,t)f(\phi)$. If we demand that the time evolution of the expectation value is driven by the generator $\frac{d}{dt}\int Pf=\int P\,Lf$, we can in turn use integration by parts to transfer the derivatives of $f$ onto $P$. For the diffusion part we have
\begin{align}
\int d\phi\,P\,\big[bf'+f''\big]
=\int d\phi\,f\,\Big[-\partial_\phi(bP)+\partial_\phi^2P\Big].
\end{align}
Boundary terms have been neglected, assuming that the distribution falls off quickly enough towards infinity.

Now for the jump part we wish to similarly move $f$ to the front of the integrand. To this end we can exploit the translation invariance of the integral over $\phi$ to shift $\phi\to\phi-\eta$
\begin{align}
&\lambda_0\!\int d\phi\,P(\phi)\!\int d\eta\,q(\eta|\phi)f(\phi+\eta)\\
&=\lambda_0\!\int d\phi\,f(\phi)\!\int d\eta\,q(\eta|\phi-\eta),
P(\phi-\eta)
\end{align}
while the second term was simply an expectation value to begin with
\begin{align}
-\lambda_0\!\int d\phi\,P(\phi)f(\phi)\!\int d\eta\,q(\eta|\phi)
&=-\lambda_0\!\int d\phi\,f(\phi)\,P(\phi).
\end{align}
Let us define the jump-rate density
\begin{align}
W(\phi'|\phi)\;\equiv\;\lambda_0\;q(\phi'-\phi\,|\,\phi),
\label{eq:Wdef}
\end{align}
and rename the integration variable $\phi'=\phi-\eta$, which reveals an intuitive gain-loss structure. Since we chose the function $f$ arbitrarily, the relations hold pointwise for the probability density
\begin{widetext}
\begin{align}
\partial_t P=\underbracket{\partial_\phi\!\left[-b\,P+\partial_\phi
P\right]}_{\text{Fokker--Planck}}
+\underbracket{\int d\phi'\left[W(\phi|\phi')P(\phi')-W(\phi'|\phi)P(\phi)
\right]}_{\text{master equation (jumps)}} .
\label{eq:fpme}
\end{align}    
\end{widetext}
This generalized Fokker-Planck equation has the CK form and furnishes the basis to construct jumps that allow us to sample from the desired distribution. 

Since we know from standard stochastic quantization that for $b=-S'$ the diffusive part of \cref{eq:fpme} annihilates $P\propto {\rm exp}[-S]$, we must only make sure that the jump term also vanishes in that case. I.e. 
\begin{align}
\int d\phi'\left[W(\phi|\phi')\,e^{-S(\phi')}-W(\phi'|\phi)\,e^{-S(\phi)}
\right]=0\quad\text{for all }\phi .
\label{eq:global}
\end{align}
Following the conventional construction of Markov-chain Monte Carlo methods (see e.g. \cite{Sokal:1996cargese}) we may achieve this result by imposing detailed balance
\begin{align}
W(\phi'|\phi)\,e^{-S(\phi)}=W(\phi|\phi')\,e^{-S(\phi')} .
\label{eq:db}
\end{align}

\subsection{Jump design}
\label{sec:jumpd}
In this work we will consider three types of jump probabilities, offering different levels of freedom to attack ergodicity in the presence of barriers.

\textbf{\textit{(A) Additive proposals}}: We choose a fixed and even density $q_0(\eta)=q_0(-\eta)$ and assure detailed balance via the Metropolis strategy
\begin{align}
W(\phi'|\phi)=\lambda_0\;q_0(\phi'-\phi)\;
\min\!\big(1,\,e^{-[S(\phi')-S(\phi)]}\big).
\label{eq:WA}
\end{align}
It is straightforward to verify this choice. Since
$\min(1,e^{-\Delta})\,e^{-S(\phi)}=\min\!\big(e^{-S(\phi)},
e^{-S(\phi')}\big)$, we get
\begin{widetext}
\begin{align}
W(\phi'|\phi)\,e^{-S(\phi)}
=\lambda_0\,q_0(\phi'-\phi)\,
\min\!\big(e^{-S(\phi)},e^{-S(\phi')}\big)
=W(\phi|\phi')\,e^{-S(\phi')},
\label{eq:dbcheck}
\end{align}
\end{widetext}

\textbf{\textit{(B) Invertible maps}}: The fact that we can now change the field by a finite amount opens up the possibility to incorporate prior information into the update step. If we choose an invertible map $T$, we may apply it or its inverse $T^{-1}$ with equal probability, weighting again with the Metropolis factor
\begin{widetext}
\begin{align}
W(\phi'|\phi)=\frac{\lambda_0}{2}\Big[
\delta\big(\phi'-T\phi\big)\,a_{T}(\phi)
+\delta\big(\phi'-T^{-1}\phi\big)\,a_{T^{-1}}(\phi)\Big],
\qquad
a_T(\phi)=\min\!\big(1,\,|J_T(\phi)|\,e^{-[S(T\phi)-S(\phi)]}\big),
\label{eq:WB}
\end{align}
\end{widetext}
Note that due to the use of the Dirac delta function, the Jacobian of the transformation $J_T$ appears in the Metropolis term.

Let us derive this relation carefully. When substituting \cref{eq:WB} into the detailed balance relation \cref{eq:db} we find four terms. 

On the left-hand side there is $\delta(\phi'-T\phi)$, which lives on the surface $\phi'=T\phi$ and
$\delta(\phi'-T^{-1}\phi)$ on $\phi'=T^{-1}\phi$. On the right-hand side,
$\delta(\phi-T^{-1}\phi')$ is supported where $\phi'=T\phi$, the same surface as the first left-hand term. In addition $\delta(\phi-T\phi')$ is supported, where $\phi'=T^{-1}\phi$. 

To see that detailed balance is achieved on both surfaces we need to compare coefficients in front of the delta functions
\begin{widetext}
    \begin{align}
\delta\big(\phi-T^{-1}\phi'\big)
=\big|\det J_{T^{-1}}(\phi')\big|^{-1}\,\delta\big(\phi'-T\phi\big)
=\big|\det J_{T}(\phi)\big|\,\delta\big(\phi'-T\phi\big),
\label{eq:deltarule}
\end{align}
\end{widetext}
which leads us to 
\begin{widetext}
\begin{align}
\phi'=T\phi:&\qquad
a_T(\phi)\,e^{-S(\phi)}=|J_T(\phi)|\;a_{T^{-1}}(T\phi)\,e^{-S(T\phi)},
\label{eq:cond1}\\
\phi'=T^{-1}\phi:&\qquad
a_{T^{-1}}(\phi)\,e^{-S(\phi)}
=|J_{T^{-1}}(\phi)|\;a_{T}(T^{-1}\phi)\,e^{-S(T^{-1}\phi)}.
\label{eq:cond2}
\end{align}
\end{widetext}
Interestingly, these two equations \cref{eq:cond1} and \cref{eq:cond2} are not independent, since we can substitute $\phi\to T^{-1}\phi$ in \cref{eq:cond1} and use $|J_{T^{-1}}(\phi)|=|J_T(T^{-1}\phi)|^{-1}$ to reproduce \cref{eq:cond2}. 

Now we have to make sure that the Metropolis form of \cref{eq:WB} is correct. Making the definition of the $a$ terms explicit we have
\begin{widetext}
    \begin{align}
a_T(\phi)\,e^{-S(\phi)}
=\min\!\Big(1,\;|J_T(\phi)|\,e^{-[S(T\phi)-S(\phi)]}\Big)\,e^{-S(\phi)}
=\min\!\Big(e^{-S(\phi)},\;|J_T(\phi)|\,e^{-S(T\phi)}\Big).
\label{eq:lhschain}
\end{align}
\begin{align}
a_{T^{-1}}(T\phi)
=\min\!\Big(1,\;\big|J_{T^{-1}}(T\phi)\big|\;
e^{-[S(T^{-1}(T\phi))-S(T\phi)]}\Big)
=\min\!\Big(1,\;|J_T(\phi)|^{-1}\,e^{-[S(\phi)-S(T\phi)]}\Big),
\label{eq:revacc}
\end{align}
\end{widetext}
which, when used to evaluate the RHS of \cref{eq:cond1}, lead to 
\begin{widetext}
\begin{align}
|J_T(\phi)|\,e^{-S(T\phi)}\;a_{T^{-1}}(T\phi)
&=\min\!\Big(|J_T(\phi)|\,e^{-S(T\phi)},\;
|J_T(\phi)|\,e^{-S(T\phi)}\cdot|J_T(\phi)|^{-1}e^{S(T\phi)-S(\phi)}\Big)
\nonumber\\
&=\min\!\Big(|J_T(\phi)|\,e^{-S(T\phi)},\;e^{-S(\phi)}\Big).
\label{eq:rhschain}
\end{align}
\end{widetext}
This expression now agrees with the LHS of \cref{eq:cond1}, proving that detailed balance is indeed achieved.

There are two important cases that we will exploit in the following. If the map $T$ is measure preserving, i.e. $J_T=1$, then acceptance reduces to the standard Metropolis term. This includes a large class of maps, from translations of angles or real fields via site-wise unitary rotations of constrained fields to left-multiplication by a fixed group element for gauge
links, since the Haar measure is invariant. If the map is a symmetry of the action then every jump is accepted.

\textbf{\textit{(C) Families of invertible maps}}: Here we take inspiration from work originally presented in the context of discrete systems, which, thanks to the jump-diffusion framework, may now be applied also in stochastic quantization. In Ref.~\cite{Zanella:2020} state informed proposals were explored, which we here adopt as third jump construction strategy.

If we are dealing with a field of multiple degrees of freedom, e.g. distributed on a spacetime grid, there may exist a variety of transformations $T_m$ labelled by parameters $m$, which can denote their point of application, their size or orientation. Let's assume that we have a family $\{T_m\}_{m\in M}$ of measure-preserving maps, such that the inverse of every member is again a member, $T_m^{-1}=T_{m^-}$ with $m^-\in M$. Then we may select a realization of such maps with a state dependent probability $p(m|\phi)$ leading to a jump probability density
\begin{align}
W(\phi'|\phi)=\lambda_0\sum_{m\in M}p(m|\phi)\,
\delta\big(\phi'-T_m\phi\big)\,a_m(\phi),
\label{eq:WC}
\end{align}
The main difference from construction (B) is the additional probability $p(m|\phi)$, which leads to a modified detailed balance relation. For every $m$ and every $\phi$, with $\phi'=T_m\phi$, we must impose
\begin{align}
a_m(\phi)\,p(m|\phi)\,e^{-S(\phi)}
=a_{m^-}(\phi')\,p(m^-|\phi')\,e^{-S(\phi')},
\label{eq:condC}
\end{align}
which is the analog of \eqref{eq:cond1}, with the selection
probabilities now appearing alongside the acceptances.

We must still solve \cref{eq:condC}. Since the new elements are the selection probabilities let us take an educated guess for the modified Metropolis Hastings ratio
\begin{align}
\nonumber&R_m(\phi)\;\equiv\;\frac{p(m^-|\phi')\,e^{-S(\phi')}}
{p(m|\phi)\,e^{-S(\phi)}}
=\frac{p(m^-|\phi')}{p(m|\phi)}\;e^{-\Delta S_m(\phi)},\\
&\Delta S_m(\phi)=S(T_m\phi)-S(\phi),
\label{eq:Rdef}
\end{align}
and note its reciprocity: evaluating \eqref{eq:Rdef} for the
reverse move ($m\to m^-$, $\phi\to\phi'$), the inverse appears exactly as needed and similar to (B), $T_{m^-}\phi'=T_m^{-1}T_m\phi=\phi$, so numerator and
denominator exchange roles and
\begin{align}
R_{m^-}(\phi')=\frac{p(m|\phi)\,e^{-S(\phi)}}{p(m^-|\phi')\,e^{-S(\phi')}}
=\frac{1}{R_m(\phi)}.
\label{eq:recip}
\end{align}
If we use this Metropolis Hastings acceptance
\begin{align}
a_m(\phi)=\min\!\big(1,\,R_m(\phi)\big),
\label{eq:MH}
\end{align}
we can show that \cref{eq:condC} is fulfilled
\begin{widetext}
\begin{align}
a_m(\phi)\,p(m|\phi)\,e^{-S(\phi)}
=R_m(\phi)\min\!\big(1,R_m(\phi)^{-1}\big)\,p(m|\phi)\,e^{-S(\phi)}
=\min\!\big(1,R_{m^-}(\phi')\big)\,p(m^-|\phi')\,e^{-S(\phi')},
\label{eq:MHcheck}
\end{align}
\end{widetext}
Now correct convergence is guaranteed for any selection probability and we may try to find the optimal choice for $p$.

The strategy we chose among those discussed in \cite{Zanella:2020} is to balance two extreme cases. Let's consider a selection probability as a function of the probability induced by the change in action $p(m|\phi)\propto
g\big(e^{-\Delta S_m(\phi)}\big)$. The one extreme is the uniform, i.e. uninformed, choice, where $g=1$. Plugging into the Metropolis Hastings term \cref{eq:Rdef} we find that $R_m=e^{-\Delta S_m}$. Here the full cost of the move appears in the accept-reject step.

The other extreme is the choice of $g(x)=x$, where $p\propto e^{-\Delta S_m}$, which when substituting gives
\begin{align}
R_m(\phi)=\frac{e^{+\Delta S_m}/\tilde Z(\phi')}
{e^{-\Delta S_m}/\tilde Z(\phi)}\;e^{-\Delta S_m}
=\frac{\tilde Z(\phi)}{\tilde Z(\phi')}\;e^{+\Delta S_m}:
\end{align}
Now the cost $\Delta S_m$ enters with the opposite sign. For the forward probability we select an easy move but then the reverse is highly unlikely, as it competes against all other attractive moves available for $\phi'$. 

If on the other hand we take the midpoint $g(x)=\sqrt{x}$ the burden of cost is shared by the forward and backward move and the only remnant in the the Metropolis Hastings ratio is the ratio between how many possibilities for jumps there are forward vs. backward. I.e. the locally balanced weight is
\begin{widetext}
\begin{align}
p(m|\phi)=\frac{e^{-\Delta S_m(\phi)/2}}{Z(\phi)},\qquad
Z(\phi)=\sum_{m\in M}e^{-\Delta S_m(\phi)/2},
\label{eq:lbweight}
\end{align}
\end{widetext}
With this balanced choice, the acceptance simplifies considerably. The cost for a reverse jump with
$\phi'=T_m\phi$ and $T_{m^-}\phi'=\phi$ is now
\begin{align}
\Delta S_{m^-}(\phi')&=S\big(T_{m^-}\phi'\big)-S(\phi')\\
&=S(\phi)-S(T_m\phi)=-\Delta S_m(\phi).
\label{eq:dsrecip}
\end{align}
What is then the probability for the reverse selection, when evaluated on the proposed configuration? The answer is
\begin{align}
p(m^-|\phi')=\frac{e^{-\Delta S_{m^-}(\phi')/2}}{Z(\phi')}
=\frac{e^{+\Delta S_m(\phi)/2}}{Z(\phi')}.
\label{eq:revweight}
\end{align}
With these expressions at hand let us insert \cref{eq:lbweight} and \cref{eq:revweight} into the
ratio \cref{eq:Rdef} for which we obtain
\begin{align}
R_m(\phi)&=\frac{e^{+\Delta S_m(\phi)/2}\big/Z(\phi')}
{e^{-\Delta S_m(\phi)/2}\big/Z(\phi)}\;e^{-\Delta S_m(\phi)}\\
&=\frac{Z(\phi)}{Z(\phi')}\; e^{\frac{\Delta S_m}{2}+\frac{\Delta S_m}{2}-\Delta S_m}=\frac{Z(\phi)}{Z(\phi')},
\end{align}
so that
\begin{align}
a_m(\phi)=\min\!\left(1,\;\frac{Z(\phi)}{Z(\phi')}\right).
\label{eq:zanella}
\end{align}
The dependence on which parameter $m$ was chosen has disappeared entirely from the acceptance. 

\subsection{Analyzing jump efficiency}
\label{sec:analyzej}
For the large class of measure-preserving symmetric jumps that we have discussed in construction principle (B), we can establish an identity for the acceptance that can be used to rank the efficiency of the construction a priori. Here we follow the strategy deployed in the literature to investigate the acceptance rates of the HMC algorithm \cite{Duane:1987de,Creutz:1988wv,Gupta:1988js,Kennedy:1991nn,Kennedy:2000ju}. Let us consider the equilibrium distribution of the jump cost, where we integrate over all proposals. Here it does not matter whether the proposals are ultimately accepted or not.
\begin{widetext}
\begin{align}
\big\langle e^{-\Delta S}\big\rangle_{\rm eq}
=\int d\phi\,\frac{e^{-S(\phi)}}{Z}\sum_{\pm}\frac12\,e^{-[S(T^{\pm}\phi)-S(\phi)]}=\sum_{\pm}\frac12\int d\phi\,\frac{e^{-S(T^{\pm}\phi)}}{Z}=1,
\label{eq:work}
\end{align}
\end{widetext}
where we have used in the last equality that we can change variables $\phi\to T^{\pm}\phi$ with unit Jacobian.

We may exploit this relation in the following fashion. Let us assume that a single update $\Delta S$ arises from multiple changes to weakly correlated degrees of freedom. Invoking the central limit theorem, the distribution of $\Delta S\sim{\cal N}(\mu,\sigma)$ can be approximated as Gaussian. Using completion of the square one establishes that $\langle e^{-\Delta S}\rangle=e^{-\mu+\sigma^2/2}$. Together with our relation $\langle e^{-\Delta S}\big\rangle=1$ this forces 
\begin{align}
    \mu=\sigma^2/2\label{eq:constr}
\end{align}

Now we can turn to the acceptance itself. The Metropolis step splits into two contributions, increasing or decreasing the action value
\begin{align}
\big\langle\min(1,e^{-\Delta S})\big\rangle
=\underbracket{P(\Delta S\le0)}_{\text{always accepted}}
\;+\;\underbracket{\big\langle e^{-\Delta S}\,
\mathbf 1_{\Delta S>0}\big\rangle}_{\text{accepted with weight}} .
\end{align}

The first contribution we can understand by making the fluctuating variable explicit $\Delta S= \mu+\sigma Z$ with $Z\sim{\cal N}(0,1)$. Then 
\begin{align}
    P(\mu+\sigma Z\le0)=P(Z<-\mu/\sigma)=\Phi(-\mu/\sigma)=\Phi(-\sigma/2)
\end{align}
where in the last step we have inserted the constraint \cref{eq:constr}. Here $\Phi=\frac{1}{2}(1+{\rm erf}(x/\sqrt{2}))$ refers to the cummulative distribution function of the normal distribution. Since we are evaluating for negative argument we use  $\Phi(-x)=\frac{1}{2}{\rm erfc}(x/\sqrt{2})$. For the second contribution we must evaluate the weight factor 
\begin{widetext}
\begin{align}
\big\langle e^{-\Delta S}\,\mathbf 1_{\Delta S>0}\big\rangle
=\int_0^\infty\!\! dx\;e^{-x}\,\mathcal N(x;\mu,\sigma)
=e^{-\mu+\sigma^2/2}\!\int_0^\infty\!\! dx\;
\mathcal N(x;\mu-\sigma^2,\sigma)
=e^{-\mu+\sigma^2/2}\;\Phi\!\Big(\frac{\mu-\sigma^2}{\sigma}\Big).
\end{align}
\end{widetext}
Together with the constraint \cref{eq:constr} this term too becomes $\Phi(-\sigma/2)$. Combining the two terms we arrive at the acceptance estimate
\begin{align}
    {\rm acceptance}=2\Phi(-\sigma/2)={\rm erfc}(\sigma/\sqrt{8})\label{eq:acceptpred}
\end{align}
Switching mean and variance via \cref{eq:constr} one arrives at the equivalent ${\rm erfc}(\sqrt{\mu}/2)$, which is well known formula for Hybrid Monte Carlo, the classic acceptance formula in terms of the mean energy violation \cite{Kennedy:1991nn}. Our derivation shows that it arises generally from measure-preserving symmetric proposals, of which the HMC
trajectory is but one example.

The acceptance estimate must be modified when we instead use the informed updates of construction (C). The key feature of the state informed jumps is that we are able to replace the standard Metropolis update by $\min(1,Z/Z')$, so that the corresponding acceptance cost variable is
\begin{align}
    \Delta S_{\rm eff}=\ln\!\big(Z(\phi')/Z(\phi)\big)
\end{align}
As long as we replace $\Delta S$ by $\Delta S_{\rm eff}$ all other steps of the derivation above follow if we start by averaging over all fields $\phi$ and parameters $m$ 
\begin{widetext}
\begin{align}
\big\langle e^{-\Delta S_{\rm eff}}\big\rangle
=\sum_m\int d\phi\;\frac{e^{-S(\phi)}}{Z_S}\,p(m|\phi)\,\frac{p(m^-|T_m\phi)\,e^{-S(T_m\phi)}}{p(m|\phi)\,e^{-S(\phi)}}
=\sum_m\int d\phi'\;p(m^-|\phi')\,\frac{e^{-S(\phi')}}{Z_S}
=1.
\label{eq:workC}
\end{align}
\end{widetext}
I.e. the acceptance $\text{acceptance}=\mathrm{erfc}\big(\sigma_{\rm eff}/\sqrt8\big)$
with $\sigma_{\rm eff}=\mathrm{sd}\big(\ln(Z'/Z)\big)$.

The estimates derived above tell us that the design of jumps is in essence a variance minimization problem for $\Delta S$ or $\Delta S_{\rm eff}$. We may ask whether it is possible to test the efficacy of a jump map a priori. 

Let us assume that we are in possession of a chain of configurations that sample $e^{-S}$ or one subsector. The latter is feasible even with the standard Langevin diffusive update. Then computing $\sigma(\Delta S_{\rm eff})$ on that chain provides a relevant measure of jump efficiency away from the sector the conventional simulation was stuck in. In order to algorithmically and exhaustively explore possible jump designs with e.g. methods of machine learning, such an a priori test of efficacy will serve an important role.

\section{Applications}
\label{sec:application}

In this section we apply the jump construction strategies (A)-(C) to simple model systems. The code and data analysis scripts used to generate the data and figures are available open access at the Zenodo repository \cite{rothkopf:2026}.

\subsection{Tilted double well}
\label{sec:tdw}
\begin{figure*}
    \centering
    \includegraphics[width=0.45\linewidth]{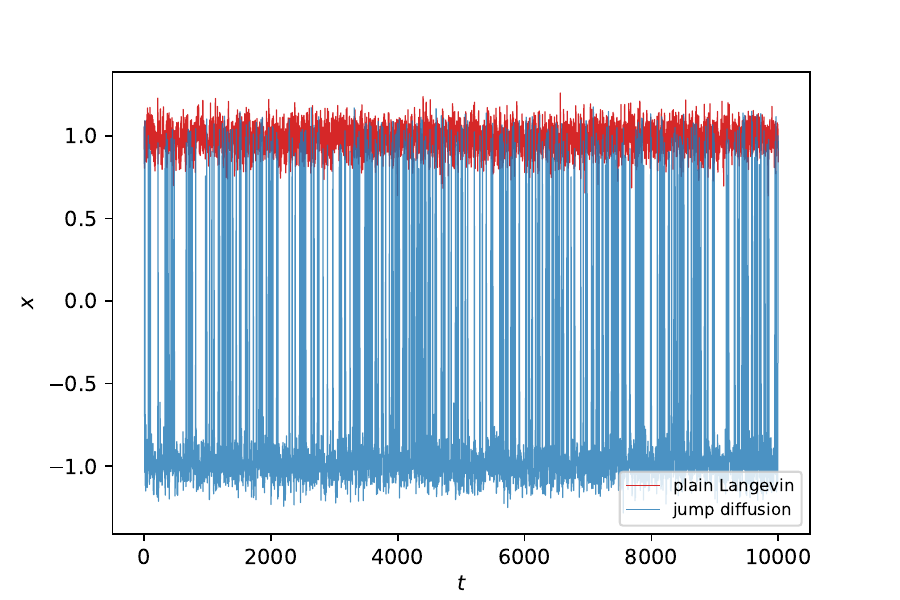}
    \includegraphics[width=0.45\linewidth]{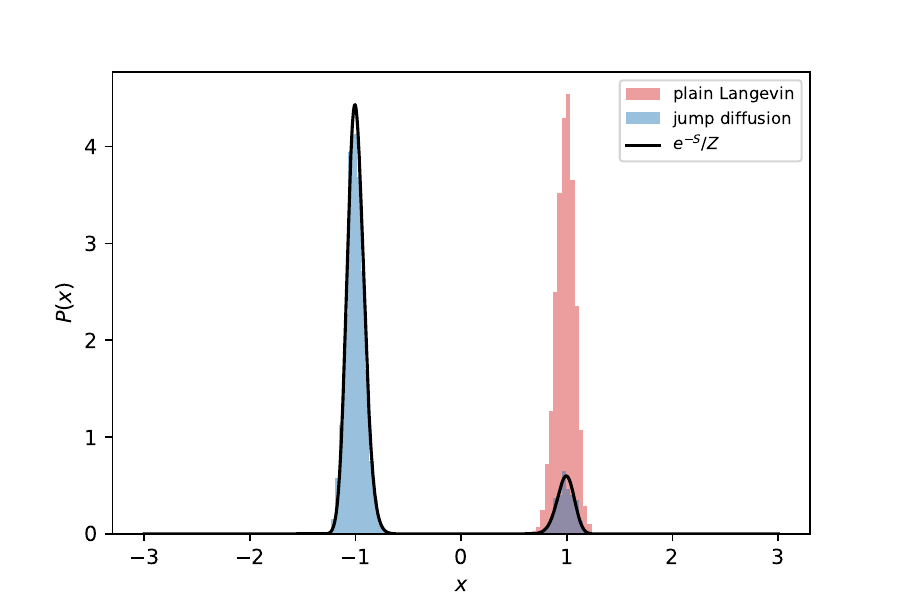}
    \caption{(left) Individual trajectory of the stochastic process sampling the tilted double well using diffusion Langevin (red) or our novel jump-diffusion approach (blue), both initialized at $x=1$. Note the frequent visits of the opposite local minima by the latter. (right) Histogram of the sampled degree of freedom for diffusive langevin (red) and the ergodical sampling by jump-diffusion (blue)}
    \label{fig:0dHist}
\end{figure*}

The simplest model to consider in which diffusive stochastic quantization easily loses ergodicity is the tilted double well. Take as action of the system $S(x)=\gamma(x^2-1)^2+cx$. If the barrier is chosen large enough a sampler placed initially in either of the two local minima remains stuck.

We set out to construct efficient jumps to restore the ergodicity and prior knowledge of the potential supplies the necessary guiding principle. In case of no tilt, the system has exact $\mathbb{Z}_2$ symmetry, a flip $x\to-x$ will carry the system from the neighborhood of one minimum to a neighborhood of the other minimum. I.e. we will land in a region with the same action. Now in the presence of a tilt this symmetry is not exact and we may over- or undershoot the neighboring region by a naive flip. Hence let us add a bit of Gaussian noise to the flip to effectively smear out the flip.

Such a proposal is thus generated in two stages: a deterministic flip $\psi=T\phi=-\phi$, whose density is a delta function
$\delta(\psi-T\phi)$, followed by a random displacement $\phi'=\psi+\xi$ with $\xi$ drawn from a Gaussian density $q_0$ of width $s$. At first sight it appears therefore to be a combination of our construction principle (A) and (B).

The density of the composite proposal however follows by marginalizing over the intermediate stage, and that convolution gets rid of the delta
\begin{align}
q(\phi'|\phi)&=\int d\psi\;\delta\big(\psi-T\phi\big)\,q_0(\phi'-\psi)=q_0\big(\phi'-T\phi\big)\\
&=\frac{1}{\sqrt{2\pi s^2}}\;e^{-(\phi'+\phi)^2/2s^2}.
\label{eq:composite}
\end{align}
The last line of \cref{eq:composite} shows that since the
argument $(\phi'+\phi)^2$ is invariant under
$\phi\leftrightarrow\phi'$, so symmetry holds for \emph{any} $q_0$, even an asymmetric one. I.e. this update can be regarded as an update of class (A) alone.

After that proposal we accept or reject with the Metropolis step to ensure detailed balance according to \cref{eq:WA}.

We demonstrate the improvement in sampling using the parameters $\gamma=20$ and $c=1$ for the tilted potential well. As seen in the left panel of \cref{fig:0dHist} as black solid line, the probability distribution ${\rm exp}[-S]$ is bimodal with no significant overlap of the two modes.

We simulate a realization of the system under discretized Langevin time with $\Delta t=10^{-3}$ up to a total sampling time of $t_{\rm max}=10^4$. The jump attempt rate is $\lambda_0=1$. We draw the random number $\xi$ from a normal distribution with width $\sigma_\xi=0.3$ chosen small compared to the distance between the two modes.

If we initialize the standard diffusion Langevin dynamics in the local extremum at $x=1$, the process will remain trapped there for all time as seen in the red curve in the left panel of \cref{fig:0dHist}. The corresponding histogram in the right panel reflects sampling only in the shallow minimum. The correct average is not recovered
\begin{align}
\langle x \rangle_{\rm true}=-0.765, \qquad \langle x \rangle_{\rm diffusion}=0.985(2)
\end{align}

On the other hand the jump-diffusion evolution shows frequent visits to regions close to the local minima in the left panel of \cref{fig:0dHist}. In this particular run 644 crossings were made. In addition to the presence of jumps we must make sure that they are balanced such that the correct relative weights of the two modes in the distribution are recovered. The blue historgram in the right panel of \cref{fig:0dHist} suggeststs that our prescription succeeds qualitatively. And indeed, the correct expectation value is recovered within uncertainty 
\begin{align}
\langle x \rangle_{\rm jumps}=-0.755(12).
\end{align}

\subsection{2d U(1) gauge theory}
\label{sec:2dU1}
While the double well illustrates that jump-diffusion in principle can improve ergodicity in sampling, we exploited prior knowledge that in more realistic theories may not be available.

To see that the freedom we describe in constructing jumps offers effective strategies also in more complicated scenarios, we turn to compact U(1) gauge theory on an $L\times L$ torus with Wilson action $S=\beta\sum_p(1-\cos\theta_p)$, where $\theta_p$ denotes the plaquette angle. The active degrees of freedom of the theory are the link angles $\theta_\mu(x)$ with $\mu\in\{0,1\}$ from which the plaquette angles are constructed via
\begin{align}
    \theta_p(x)=\theta_0(x) +\theta_1(x+\hat e_0 a) - \theta_0(x+\hat e_1 a) - \theta_1(x)
\end{align}
as finite differences between the grid nodes one lattice spacing $a$ apart.

What makes this system an ideal test case for realistic gauge theories is that its configurations carry an integer topological charge
$Q=\frac{1}{2\pi}\sum_p\bar\theta_p$ (principal values
$\bar\theta_p\in(-\pi,\pi]$) (for more details see \cite{Luscher:1981zq,Smit:1986fn}). The standard topological argument for 2d U(1) goes as follows. For a periodic lattice the plaquette angles always sum to zero $\sum_x \theta_p(x)=0$ because every link appears in exactly two plaquettes with opposite orientation. If one decomposes $\theta_p(x)=\theta_p^{\rm PV}(x)+2\pi n(x)$ with $n(x)\in \mathbb{Z}$, we get that $\sum_x \theta_p^{\rm PV}(x)=-2\pi \sum_x n(x) \in 2\pi \mathbb{Z}$ where $Q=\sum_x n(x)$ is the topological charge.

It is known that toward the continuum limit (large $\beta$ at fixed $L^2/\beta$), changing $Q$ requires an action cost related to pulling the plaquette angle over the top of the cosine, at cost $\sim2\beta$. In turn local
dynamics freezes into a single sector. 

The model is an ideal testbed because plaquette, charge distribution $P(Q)$, and topological susceptibility $\chi_t=\langle Q^2\rangle/V$ have been computed exactly in the discrete setting in Ref.~\cite{Bonati:2019ylr}.

With the intention to make this paper self contained, in the following we will develop three different jump strategies that exploit prior knowledge of the topology of this system. Among them we include the winding kernels that were previously explored in \cite{Albandea:2021lvq} in the context of a jump extended HMC.

\subsubsection{Global updates}
\label{sec:globu}

\begin{table*}[htbp]
  \centering
  \small
  \setlength{\tabcolsep}{4pt}
  \begin{tabular}{cc rr ccc ccc}
    \toprule
    & & \multicolumn{2}{c}{$\tau$} & \multicolumn{3}{c}{$\langle Q^2\rangle$} & \multicolumn{3}{c}{$\langle P\rangle$} \\
    \cmidrule(lr){3-4} \cmidrule(lr){5-7} \cmidrule(lr){8-10}
    $\beta$ & $L$ & plain & jump & plain & jump & exact & plain & jump & exact \\
    \midrule
     0.5 &  4 &    0.21 & 0.19 & $0.935 \pm 0.013$ & $0.955 \pm 0.012$ & 0.952 & $0.246 \pm 2\times 10^{-3}$ & $0.245 \pm 2\times 10^{-3}$ & 0.242 \\
     1.1 &  6 &    0.34 & 0.28 & $1.325 \pm 0.020$ & $1.366 \pm 0.020$ & 1.327 & $0.490 \pm 1\times 10^{-3}$ & $0.486 \pm 1\times 10^{-3}$ & 0.489 \\
     2.0 &  8 &    0.81 & 0.60 & $1.209 \pm 0.029$ & $1.273 \pm 0.027$ & 1.239 & $0.699 \pm 6\times 10^{-4}$ & $0.698 \pm 6\times 10^{-4}$ & 0.698 \\
     4.5 & 12 &   20.67 & 1.48 & $0.876 \pm 0.109$ & $0.965 \pm 0.040$ & 0.940 & $0.880 \pm 1\times 10^{-4}$ & $0.880 \pm 1\times 10^{-4}$ & 0.880 \\
     8.0 & 16 &  325.33 & 1.41 & 1.628\,(frozen)   & $0.846 \pm 0.036$ & 0.870 & $0.935 \pm 4\times 10^{-5}$ & $0.935 \pm 4\times 10^{-5}$ & 0.935 \\
    12.5 & 20 & $\geq 2495.00$& 1.52 & 0.000\,(frozen) & $0.813 \pm 0.031$ & 0.846 & $0.959 \pm 2\times 10^{-5}$ & $0.959 \pm 2\times 10^{-5}$ & 0.959 \\
    \bottomrule
  \end{tabular}
  \caption{Comparison of plain Langevin and jump-diffusion sampled observables and their exact values from the literature. Shown are the recorded values for topological charge fluctuations $\langle Q^2\rangle$, the autocorrelation time of $Q$ itself and the average plaquette $\langle P\rangle$. Errors encompass autocorrelation corrected statistical errors only. The finite Langevin step size of $\Delta t=2\times10^{-4}$ leads to an inherent systematic uncertainty of the same order. Hence comparisons to exact values are truncated at that order. We declare topological charge frozen if fewer than 20 transitions occur throughout the simulation, preventing a reliable autocorrelation estimate.}
  \label{tab:plain-vs-jump}
\end{table*}

\begin{figure*}
    \centering
    \includegraphics[width=0.45\linewidth]{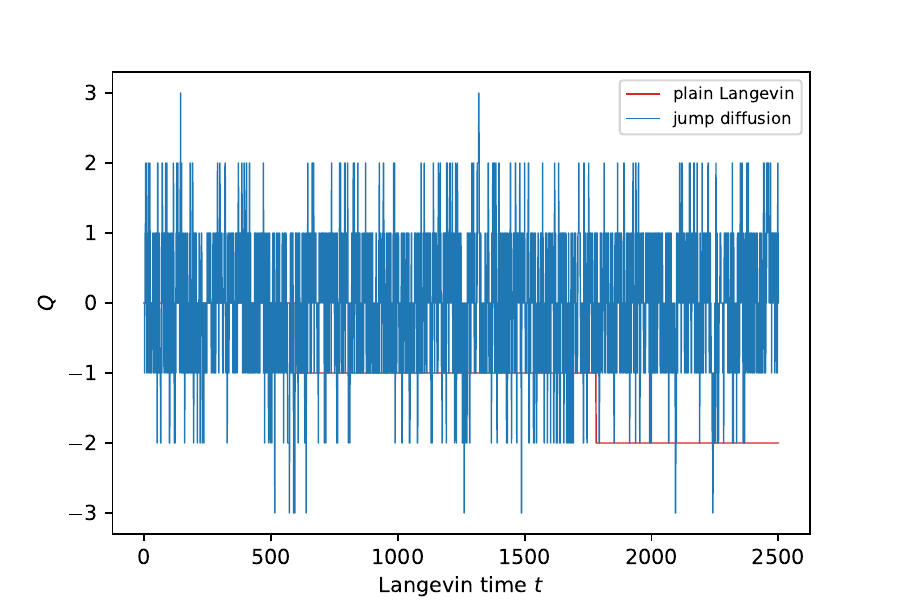}
    \includegraphics[width=0.45\linewidth]{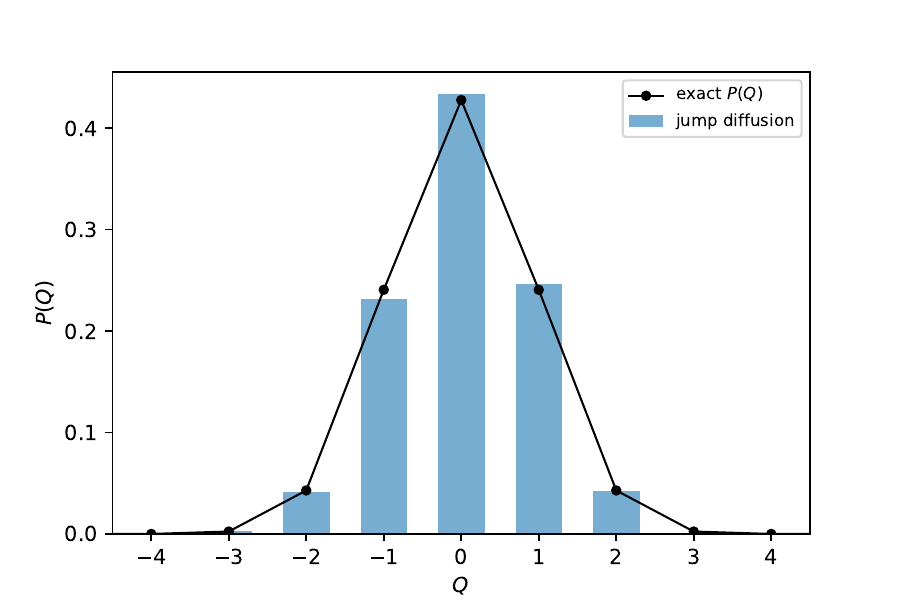}
    \caption{Visualizations at $\beta=8$, $L=16$: (left) topological charge and its transitions from plain Langevin (red solid) and the jump-diffusion update (blue). (right) Histogram of the topological charge as sampled with jump-diffusion (blue bars) compared to the exact result from the literature.}
    \label{fig:2dHist}
\end{figure*}

\begin{figure*}
    \centering
    \includegraphics[width=0.5\linewidth]{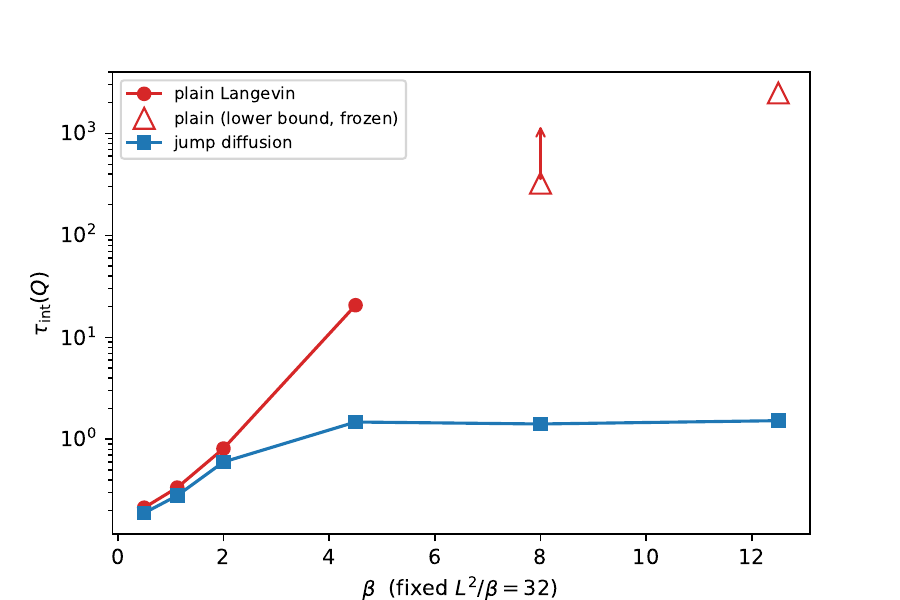}
    \includegraphics[width=0.45\linewidth]{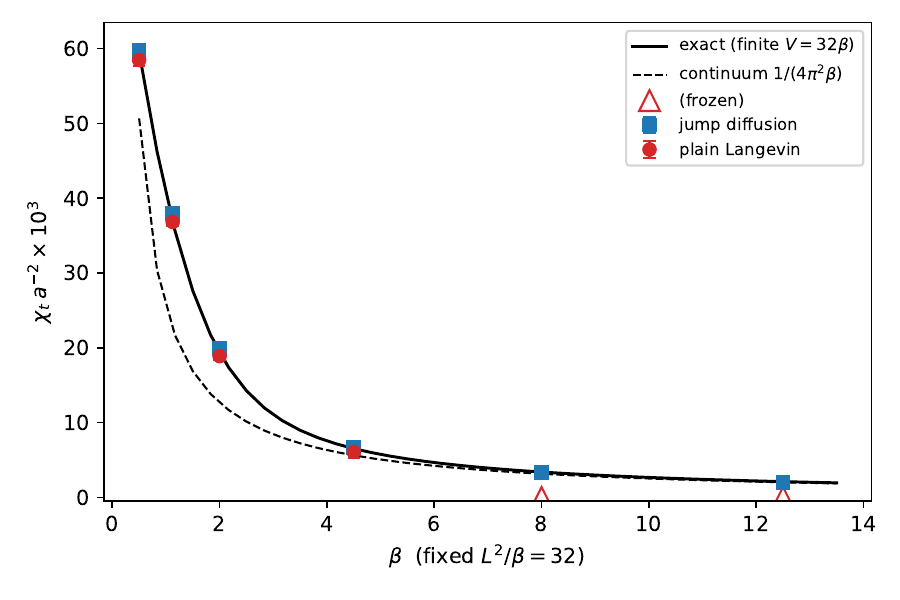}
    \caption{(left) Topological charge autocorrelation time at various lattice couplings $\beta$, as measured for plain Langevin updates (red data) and the jump-diffusion updates (blue data). Where fewer than 20 transitions occur the estimate of autocorrelation time becomes unreliable we give its value as lower bound (open triangles). (right) The values of the topological susceptibility estimated at various $\beta$ values from plain Langevin (red) and the jump-diffusion sampler (blue). The exact lattice result is given as black solid line, the continuum expression as dashed black line.}
    \label{fig:2dAutoCorr}
\end{figure*}

Since every jump undergoes a Metropolis accept-reject step we wish to make sure that it incurs the smallest possible cost. This cost is determined by the change the jump induces in the action $\beta\sum_x\big(1-\cos \theta_p(x)\big)$, which approximately reads $\frac{\beta}{2}\sum_x\theta_p(x)^2$ for small fluxes.

Just as we intended to move from one local extremum of the double well to the other, we set out to construct a jump that moves the system by a single unit of topological charge. The foal is to obtain the minimal $\sum_x\theta_p(x)^2$ under the constraint that $\sum_x\theta_p(x)=2\pi$, i.e. that it carries $|Q|=1$. Using a Lagrange multiplier we find that the optimal solution is simply the constant $\theta_p(x) = 2\pi/V$, which incurs a classical action cost of $S_{\rm class}\approx\beta V\cdot\frac12\Big(\frac{2\pi}{V}\Big)^2
=\frac{2\pi^2\beta}{V}$. In this case we are spreading the charge as thinly as possible over all plaquettes with an action cost that vanishes in the thermodynamic limit (see also \cite{Sachs:1991en}). Note that if we approach the continuum limit of $\beta\to\infty$ while keeping $\beta/V = \beta/L^2={\rm const}$, also the cost remains constant.

Now we need to find a link configuration that realizes this plaquette setting. To this end we revisit classic ideas from the literature (see e.g. \cite{Smit:1986fn}). It is understood that building such a configuration from constant raw link angles will fail, since for a periodic configuration the angles always sum to zero. But we can build a configuration using \begin{align}
\theta_p(x)=\frac{2\pi}{V}-2\pi\,\delta_{x,x_*}
\label{eq:string}
\end{align}
where one plaquette at $x_*$ reabsorbs the accumulated charge.
To make this construction concrete we exploit the fact that a plaquette angle is obtained from the link angles by finite differences. If we thus start with the ansatz
\begin{align}
    A_0(x_1,x_2)=(2\pi/V) (x_1/a), \quad A_1(x_1,x_2)=0
\end{align}
then each bulk plaquette receives $A_0(x_1+a,x_2)-A_0(x_1,x_2)=2\pi/L$, except in the last column where the wrapping around the periodic boundary leads to a remnant of plaquette angles of $-(2\pi/V)(L-a)$. This difference is taken car of by prescribing finite values to the links in $1$ direction on the final column 
\begin{align}
   A_1(L-a,x_1)=-2(\pi/L)(x_1/a)
\end{align}
These achieve the constant flux everywhere except in the corner plaquette at $(L-a,L-a)$ where a full $2\pi$ is added.

In the new jump-diffusion framework, we may add this finite global configuration to our set of gauge link angles as part of an update step. We do so by randomly shifting the link angles $\theta_i$ by $A_i$, selecting with equal probability whether to add or subtract $A_i$. Including the necessary Metropolis step this amounts to an update transition probability of
\begin{widetext}
\begin{align}
W(\theta'|\theta)=\lambda_0\left[\tfrac12\,\delta\!\big(\theta'
-\theta-A\big)+\tfrac12\,\delta\!\big(\theta'-\theta+A\big)\right]
\min\!\big(1,e^{-[S(\theta')-S(\theta)]}\big).
\label{eq:u1kernel}
\end{align}
\end{widetext}

We simulate the system on a line of constant physics $L^2/\beta=32$ using a diffusion time step of $\Delta t=2\times 10^{-4}$. The arbitrarily chosen constant jump rate is $\lambda_0=2$. The jump-diffusion process is evolved until total $t_{\rm max}=2500$ with thermalization steps discarded until $t_{\rm th}=5$.

In each step, on top of the diffusive update, we query whether the system will carry out a jump according to $\lambda_0$. Here we use the maximally spread flux configuration $A_i$, proposing it with randomly chosen sign followed by the accept-reject step. As a comparison we run the simulation with the same system parameters $\beta$ and $V$ but in the absence of jumps. 

In \cref{tab:plain-vs-jump} we collect the results for three characteristic observables comparing plain, diffusive updates, and our jump-diffusion sampler. The average plaquette $\langle P\rangle =\langle (1/V) \sum_p {\rm cos}[\theta_p]\rangle$ represents a highly local quantity. On the other hand topological charge fluctuations are encoded in $\langle Q^2 \rangle$ and are susceptible to the long range properties of the gauge fields. We also list the autocorrelation time $\tau$ estimated from the topological charge $Q$ itself. Those quantitative results are amended by explicit visualizations at $\beta=8$ in \cref{fig:2dHist} and for various $\beta$ values in \cref{fig:2dAutoCorr}.

Note that due to the finite Langevin time step $\Delta t\sim{\cal O}(10^{-4})$ and the absence of an accept-reject in the diffusion sector, the results carry an intrinsic systematic uncertainty of the same order. Any comparison to exact results is thus truncated at that order. The restoration of ergodicity we study here is not affected by this uncertainty, which can be systematically reduced via $\Delta t$ refinement.

Similar to the observations made in various studies in the past, (see, e.g., \cite{Albandea:2021lvq}) we find that a sampler with continuous trajectories, such as plain Langevin or similarly Hybrid Monte Carlo, experiences exponentially increasing autocorrelation times in the topological charge as one approaches the continuum limit. We find that at $\beta=8$ only two transitions occur in the whole of the simulation (see the red solid line in the left panel of \cref{fig:2dHist}) and at $\beta=12.5$ no transition occurs at all. In case that fewer than 20 transitions occur we declare the topological charge as frozen. Similarly the autocorrelation corrected uncertainty of $\langle Q^2\rangle$ grows in the plain Langevin updates.

At the same time we find that, as expected, the fully smeared flux jump update manages to maintain an efficient transition rate between topological sectors, even when the plain diffusive updates is already frozen. The blue solid line in the left panel of \cref{fig:2dHist} reflects this fact. In the right panel of \cref{fig:2dHist} we compare the normalized histogram of our jump-diffusion sampled topological charge (blue bars) with the exact solution from \cite{Bonati:2019ylr} and we find excellent agreement.

The exponential increase of the autocorrelation time in plain Langevin sampling is visualized in the left panel of \cref{fig:2dAutoCorr} via the red data points. Where fewer than 20 transitions occurred we the autocorrelation estimate can only be taken as lower bound and is indicated via an open triangle. The jump-diffusion updates asymptote at an almost constant value instead of showing a marked increase. 

The right panel of \cref{fig:2dAutoCorr} compares the value for the topological susceptibility $\chi_t=\langle Q^2\rangle/V$ from the plain Langevin updates (red data) to those from the jump-diffusion sampler (blue data). From the literature we take the exact values for the discretized lattice theory (black solid) and the continuum values (black dashed). Note that as the two curves become indistinguishable close to $\beta=8$. It is here that the plain Langevin updates become frozen (open triangles).

\subsubsection{Local updates}
\label{sec:locu}

\begin{table*}[htbp]
  \centering
  \small
  \setlength{\tabcolsep}{4pt}
  \begin{tabular}{c rr ccc ccc}
    \toprule
    & \multicolumn{2}{c}{$\tau_{\mathrm{wnd}}$} & \multicolumn{3}{c}{$\langle Q^2\rangle$}
      & \multicolumn{3}{c}{$\langle P\rangle$} \\
    \cmidrule(lr){2-3} \cmidrule(lr){4-6} \cmidrule(lr){7-9}
    $L_w$ & uninformed & informed & uninformed & informed & exact & uninformed & informed & exact \\
    \midrule
    12 &   3.66 &  --- & $0.829 \pm 0.046$ & ---               & 0.870 & $0.935 \pm 4\times10^{-5}$ & ---                            & 0.935 \\
    10 &   4.10 &  --- & $0.883 \pm 0.053$ & ---               & 0.870 & $0.935 \pm 4\times10^{-5}$ & ---                            & 0.935 \\
     8 &   8.18 & 0.99 & $0.949 \pm 0.079$ & $0.835 \pm 0.025$ & 0.870 & $0.935 \pm 4\times10^{-5}$ & $0.935 \pm 4\times10^{-5}$ & 0.935 \\
     6 &   8.24 & 1.10 & $0.782 \pm 0.074$ & $0.847 \pm 0.028$ & 0.870 & $0.935 \pm 4\times10^{-5}$ & $0.935 \pm 4\times10^{-5}$ & 0.935 \\
     4 &  15.35 & 1.05 & $0.764 \pm 0.100$ & $0.870 \pm 0.026$ & 0.870 & $0.935 \pm 4\times10^{-5}$ & $0.935 \pm 4\times10^{-5}$ & 0.935 \\
     2 & 243.94 & 2.11 & 0.624\,(frozen)   & $0.886 \pm 0.035$ & 0.870 & $0.935 \pm 4\times10^{-5}$ & $0.935 \pm 4\times10^{-5}$ & 0.935 \\
    \bottomrule
  \end{tabular}
  \caption{Winding updates at $\beta = 8.0$, $L = 16$ using both the uninformed and informed updates discussed in the main text. We record here the values for topological charge fluctuations $\langle Q^2\rangle$, the autocorrelation time of $Q$ itself and the average plaquette $\langle P\rangle$. Errors encompass autocorrelation corrected statistical errors only. The finite Langevin step size of $\Delta t=2\times10^{-4}$ leads to an inherent systematic uncertainty of the same order. Hence comparisons to exact values are truncated at that order. We declare topological charge frozen if fewer than 20 transitions occur throughout the simulation, preventing a reliable autocorrelation estimate.}
  \label{tab:wnd-combined}
\end{table*}

\begin{figure*}
    \centering
    \includegraphics[width=0.45\linewidth]{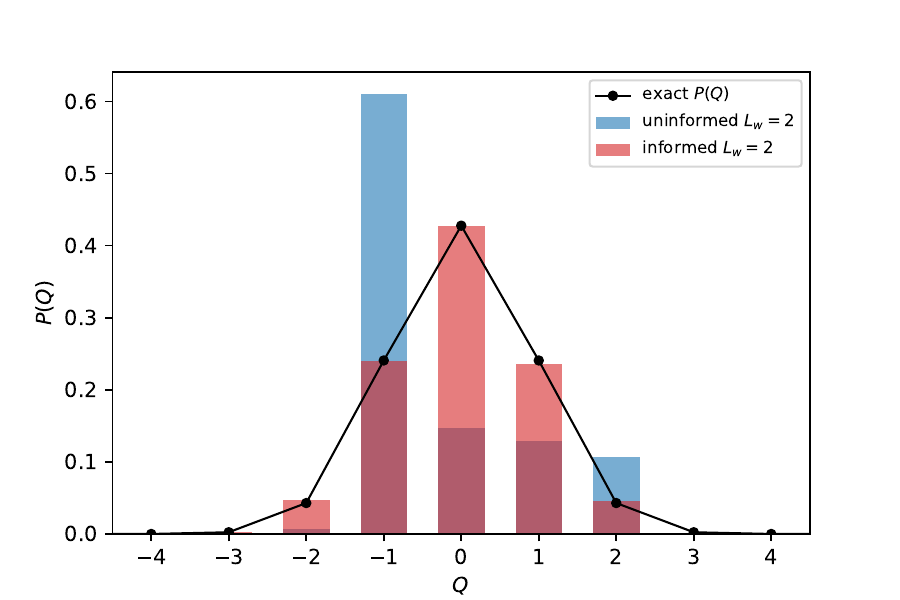}
    \includegraphics[width=0.45\linewidth]{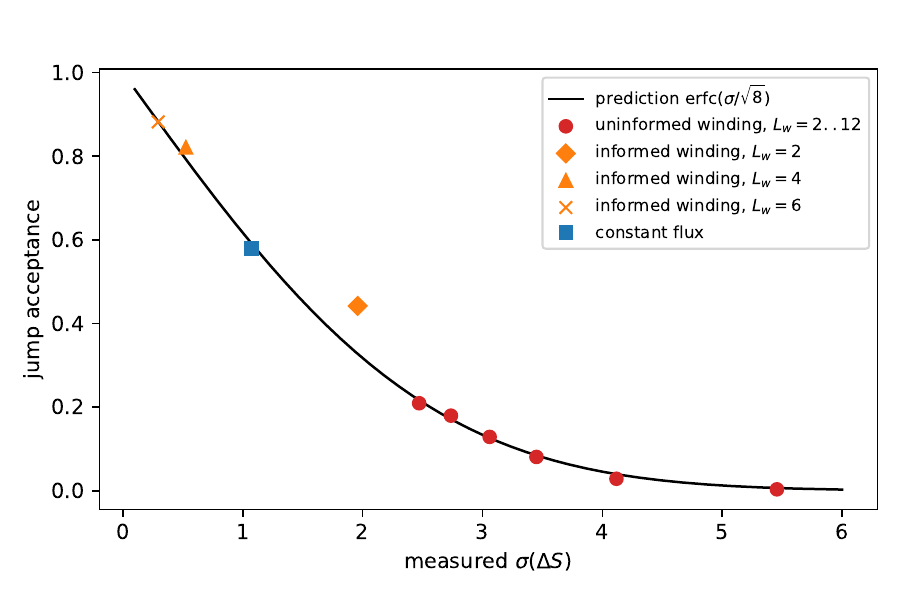}
    \caption{(left) Histogram of topological charge sampled at $\beta=8$, $L=16$ with small $L_w=2$ uninformed (A) winding jumps (blue bars) and using state informed (C) jumps (red bars). The exact result is given as black solid line. (right) Jump acceptance vs. the standard deviation of the relevant action cost. Following \cref{sec:analyzej} $\Delta S$ is used for uninformed jumps and $\Delta S_{\rm eff}$ for informed jumps. The uninformed global constant flux update (blue data) shows higher acceptance than the local uninformed winding updates (red data). The smaller the extent $L_w$, the lower the acceptance. On the other hand, informed winding updates, even for small $L_w=2$ (orange square) exceed the acceptance efficiency for the largest uninformed winding update. Informed $L_w=4$ (orange triangle) and $L_w=6$ (orange cross) updates even improve on the global constant flux update.}
    \label{fig:2dHist2}
\end{figure*}

The efficiency of the global update is a coincidence of the 2d U(1) theory, where the topological charge is linear in the field, while the action is quadratic. In realistic theories, both the action and the topological charge contain the same powers of field strength, leading to the well known lower limit on the cost for a topology changing update via the classic BPST argument $0\leq \int d^4x (F\pm \tilde F)^2 = \int d^4x( 2 F^2 \pm 2 F\tilde F) = 2(S\pm Q)$ (for details see \cite{Belavin:1975bpst,Bogomolny:1975de}). I.e. there exists a lower limit for the cost of transitions between the topological sectors.

In addition, when matter fields are part of the theory then localized updates incur extra cost due to hopping terms, which contain covariant derivatives that feel the phases of the jump map. The more delocalized the update, the higher the cost. The matter sector therefore forces us to consider localized jump updates.

To explore the efficiency of topology changing jump updates with finite extent, we turn to a class of jumps that have been investigated in the past. These winding updates introduced in the context of the jump extended Hybrid Monte Carlo scheme wHMC in \cite{Albandea:2021lvq}, can be straightforwardly incorporated in our jump-diffusion process.

Let us briefly review their construction. Starting point is to consider a gauge transformation $\theta_\mu(x)\to\theta_\mu(x)+\omega(x)-\omega(x+\hat\mu)$, applied to all links. This of course does not change any plaquette. It would be a jump with $\Delta S\equiv0$ but also changes nothing physical. The crucial point is that one restricts the gauge transformation spatially and applies it only to links whose both endpoints lie in a square region $S_w$ of side $L_w$, leaving boundary-crossing links untouched. Deep inside $S_w$ the transformation is still pure gauge and changes nothing. On the other hand the ring of plaquettes straddling $\partial S_w$, where
transformed and untransformed links meet, acquires flux. 

To make the flux we introduce topological, we must make $\omega$ wind. To this end one orders the $4L_w$ boundary sites of $S_w$ around the perimeter and sets $\omega(x_n)=\pm\pi n/(2L_w)$, so that $\omega$ advances by $2\pi$ in one closed loop. As a real function $\omega$ is discontinuous between the last site and the first, while as a U(1) element $e^{i\omega}$ it is perfectly
single-valued .

The cost follows from counting, the total flux $2\pi$ is
spread over the $\sim4L_w$ ring plaquettes, which means $|F|\approx2\pi/(4L_w)$ each, so
\begin{align}
\Delta S_{\rm class}^{\rm wind}
\approx\frac{\beta}{2}\cdot 4L_w\Big(\frac{2\pi}{4L_w}\Big)^2
=\frac{\pi^2\beta}{2L_w},
\label{eq:windcost}
\end{align}
falling like $1/L_w$. I.e., the smaller the size, the higher the cost and the less efficient at first sight the jump will be. 

If we make sure to choose the position of the winding insertion randomly the resulting transition probability reads
\begin{widetext}
\begin{align}
W(\theta'|\theta)=\lambda_0\sum_{c\in\Lambda}\sum_{s=\pm1}
\frac{1}{2L^2}\;\delta\!\big(\theta'-\theta-s\,A^{\rm w}_c\big)\,
\min\!\big(1,e^{-[S(\theta')-S(\theta)]}\big),
\qquad A^{\rm w}_{c,\mu}(x)=A^{\rm w}_\mu(x-c),
\label{eq:windkernel}
\end{align}
\end{widetext}
which again falls into the construction class (A).

We confirm the performance of these winding updates at $\beta=8$, $L=16$, by replacing the the fully spread flux update with a winding of size $L_w$, randomly selecting in each update where to position it, in accordance with \cref{eq:windkernel}. The outcome of using these \textit{uninformed} updates is tabulated in \cref{tab:wnd-combined}. 

A helpful visualization may be found in the right panel of \cref{fig:2dHist2}. Here we plot the jump acceptance rate of our jump-diffusion simulations at $\beta=8$ against the standard deviation of $\Delta S$, the quantity we identified in \cref{eq:acceptpred}. Comparing in the \textit{uninformed} setting the maximally extended constant flux updates (blue) with the more localized winding updates (red), we see that the smaller extent is directly linked to higher cost, with the smallest $L_w=2$ update lying furthest to the right. 

For the uninformed $L_w=2$ winding update the topological charge dynamics are already effectively frozen and its histogram, shown as blue bars in the left panel of \cref{fig:2dHist2}, clearly deviates from the correct result given as black solid line.

At first it may appear that we are out of luck, but our analysis in construction principle (C) promised that by an informed selection of the parameters of the jump map we may exchange the cost of a single update by the ratio of the probabilities of all different jumps forward and back. The set of parameters we are going to use here in this procedure is the placement of the localized winding map.

To this end, before the simulation, we create a list with the winding of size $L_w$ positioned at all possible grid nodes. In a higher dimensional setting one would sparsen this grid.  In each jump update, one would compute the change in action for all of these differently positioned maps and randomly select the position to jump from, according to the local cost incurred by each jump. This implements \cref{eq:lbweight}. In order to carry out an acceptance test according to \cref{eq:zanella} we compute the local cost for all jumps backward from the new configuration too. 

The results for these state \textit{informed} updates are shown in their respective columns in \cref{tab:wnd-combined}. We find that the informed update has significantly reduced the autocorrelation time for all tested winding sizes $L_w$. Most remarkably it shows an efficient reduction also for the smallest winding $L_w=2$, restoring efficient transitions between topological sectors, as evidenced by recovering the correct histogram (red bars) in the left panel of \cref{fig:2dHist2}. 

And indeed inspecting the efficiency of the \textit{informed} updates (orange data) in the right panel of \cref{fig:2dHist2}, we find that the smallest winding $L_w=2$ (orange square) now lies above the acceptance rate of the largest uninformed winding update (red) and lies already close to the otherwise optimal constant flux map (blue). The larger windings with $L_w=4$ (orange triangle) and $L_w=6$ (orange cross) are seen to surpass any of the uninformed updates.

These results bode well for the application of jump-diffusion stochastic quantization to more realistic theories, where extended updates are disfavored by matter fields and jump maps of small extent suffer from high gauge field barrier cost.

\section{Conclusion and Outlook}
\label{sec:conclo}
We have constructed the natural generalization of conventional diffusion based stochastic quantization by adding finite jumps. The new contributions to the stochastic dynamics can be captured as an additional non-local term in a generalized Fokker-Planck equation with an intuitive gain-loss structure. Careful inspection of this term allows us to construct jump updates which, on top of diffusive dynamics, converge to the correct target distribution for Euclidean field theories $e^{-S}$.

We discussed three strategies for constructing finite jumps of increasing complexity. Both (A) additive proposals with a fixed proposal density and (B) invertible maps relied fully on prior information about the system at hand to incorporate proposals that incur a minimal barrier crossing cost. On the other hand the (C) state informed update of families of invertible maps piggybacked on the fluctuations already present in the system to apply an optimal choice of jump map that does not suffer from the full cost of the jump of an individual member of the family.

Using the tilted double well as the simplest example for construction (A) we restored ergodicity and correct sampling by finite field flips. To explore the opportunities offered by all construction types (A)-(C), we also considered jump-diffusion stochastic quantization in the 2d U(1) model. Using the fully extended map of constant flux we were able to restore correct sampling close to the continuum. In anticipation of future studies of more realistic theories we also considered winding jumps of finite extent $L_w$, which, when considered as uninformed updates, lose their effectiveness for smaller and smaller extent. However by considering state informed updates we were able to show that even the smallest $L_w=2$ winding can be used for an efficient sampling with an acceptance rate close to the fully extended flux map. Larger but still small windings $L_w=4,6$ surpassed the extended flux map when applied in the state informed fashion.

Before embarking on the study of Yang-Mills theory or full QCD, the next step will be to investigate the CP(N-1) model, which incorporates both gauge and matter fields, exposing the sampler to a similarly complex setting as when sampling genuine nuclear matter.

The original motivation for this work was the question of ergodicity in complex Langevin simulations with fermions. There it is known that sampling may become non-ergodic due to singularities in the complex plane and our hypothesis is that jumps will be able to maneuver the sampler away from these exceptional points. Of course, in the presence of a complex action there is no Metropolis test at our disposal and the design of the correct jumps remains an open research question.

Jumps as part of the stochastic quantization framework also open a direct route to the treatment of discrete systems. Since diffusion does not exist on discrete state spaces, pure jumps would have to be applied. In this work we actually took inspiration from work on discrete systems when implementing construction (C).

While loss of ergodicity due to barriers is one failure mode of diffusive stochastic quantization, critical slowing down close to critical points or in frustrated geometries is another. Here we will consider in the future another freedom of stochastic quantization, the so-called kernel freedom. It has long been known \cite{Batrouni:1985jn} that introducing kernels, (in their case in the form of Fourier acceleration) can lower autocorrelation times and speed up convergence. Connecting to our earlier work on optimal and machine learned kernels \cite{Alvestad:2023jgl,Alvestad:2022abf}, we look forward to exploring how jumps and kernels can work hand in hand to achieve that goal.

With the inclusion of jumps, the stochastic quantization framework is complete in the Markovian sense and its description through a closed form generalized Fokker-Planck equation allows us to exploit this new freedom in a well controlled fashion. We hope that this formulation will contribute new impulses to the simulation of Euclidean lattice field theories and beyond.

\acknowledgements
    A.~R.~gladly acknowledges support by the National Research Foundation of Korea under grant RS-2026-25486880 "From turbulent flows to quantum fields and gravity: energy-momentum transport via symmetry-preserving dynamical coordinate maps" and by Korea University under grant K2605081 "Ab-initio lattice simulations of the real-time dynamics of non-relativistic fermions".

\section*{Author contributions}
Motivated by the loss of ergodicity in complex Langevin simulations with fermions and after study of the stochastic process literature, A.~R.~ conceived the idea of extending stochastic quantization to jump-diffusion processes. The author used Claude AI to explore the literature in adjacent research areas, in particular regarding efficient jump proposals in discrete systems, leading to Ref.~\cite{Zanella:2020} used in construction (C). The author has written the manuscript himself and all references are verified by digital object identifiers, where available.

\bibliography{JDSQ}

\end{document}